\documentclass[twoside]{article}
\usepackage{fleqn,espcrc2}
\usepackage{graphicx}

\newcommand{\AmS}{{\protect\the\textfont2
  A\kern-.1667em\lower.5ex\hbox{M}\kern-.125emS}}

\title{The low-lying Scalar Mesons
and Related Topics}

\author{T. Kunihiro\address{Department of Physics, 
        Kyoto University,
        Sakyo-ku, Kyoto, 606-8502, Japan}%
        \thanks{
This work was partially supported by a Grant-in-Aid for Scientific Research by the Ministry
of Education, Culture, Sports, Science and Technology (MEXT) of Japan (No.20540265)
 and by the Grant-in-Aid for the global COE program
``The Next Generation of Physics, Spun from Universality and Emergence''
 from MEXT.
}
, S. Muroya\address{Department of Comprehensive  Management, 
Matsumoto University, Nagano 390-1295, Japan},
A. Nakamura\address{
RIISE, Hiroshima University, Higshi-Hiroshima 739-8521, Japan},
C. Nonaka\address{Department of Physics, Nagoya University, Nagoya 464-8602, Japan},
M. Sekiguchi\address{Department of Mathematics and Science,
 Kokushikan University, Tokyo 154-8515, Japan}
and 
H. Wada\address{
Faculty of Political Science and Economics, Kokushikan University, Tokyo 154-8515, Japan}
}
       
\begin{document}

\begin{abstract}
After presenting the  motivations to explore the low-lying scalar mesons
such as the $\sigma$, $a_0$ and  $\kappa$
 in the unquenched as well as  quenched lattice QCD,
 we review  the works done by our collaboration(SCALAR Collaboration) 
with a what-to-do-next list. We briefly mention the imporatance to explore
the $N_c$ dependence of and possible effects of the $U_A(1)$ anomaly
to the properties the low-lying scalar mesons.
\end{abstract}

\maketitle

\section{Introduction}

The low-lying scalar mesons have now acquired much renewed interest since
1990's when extensive analyses claimed the existence of the $\sigma$ meson pole  
of  the $\pi$-$\pi$  $S$-matrix in the comlex energy plane in the $I=J=0$ channel.
In these analyses, 
the significance of respecting chiral symmetry, unitarity and crossing
symmetry has been recognized and emphasized 
to reproduce the phase shifts both in the $\sigma$ (s)- and $\rho$ (t)-channels
with a low-mass  $\sigma$ pole\cite{Igi:1998gn}.
One of the most elaborated analyses\cite{Caprini:2005zr}
identify the $\sigma$ pole
at  $M_{\sigma}=441- i 272$ MeV.
The low-lying $\sigma$ meson is also seen in decay processes from heavy particles\cite{Aitala:2000xu}.
Such a low-lying isoscalar and scalar meson 
has been called for and known to 
play a significant role in nuclear and hadron physics\cite{Kunihiro:2001ra}.

The significance of the $\sigma$-degrees of freedom may become
more apparent at finite temperature($T$) and/or density, where chiral restoration
is believed to occur\cite{Hatsuda:2001da}.
The order parameter of the chiral transition is the chiral condensate
$\langle \bar{\psi}\psi\rangle$.
To identify the critical temperature $T_c$ of chiral transition,
the chiral susceptibility, 
$\chi_m\equiv \frac{\partial}{\partial m}\langle \bar{\psi} \psi \rangle$
$=\langle :({\bar {\psi}} \psi)^2: \rangle$, is calculated;
the temperature at which $\chi_m$ takes the  peak is
identified as  $T_c$.
It is noteworthy that $\chi_m$ describes the
fluctuations of the order parameter in the $\sigma$ direction.
The lattice simulations\cite{Karsch:2001cy}  do show that $\chi_m$ takes a
peak structure, which means that the $\sigma$ degrees of
freedom is physically relevant around the critical point, at least. 
The development of the peak of $\chi_m$
is equivalently describable as a softening of the $\sigma$-like excitation around
 $T_c$; 
the lattice simulations show that the generalized $\sigma$ mass as defined by 
$m_{\sigma}^{\rm gen} \equiv \chi_m^{-1/2}$
takes the minimum around $T_c$.
One should notice, however,
that $m_{\sigma}^{\rm gen}$ is
not the dynamical mass defined as  a pole of the time-correlator.

So far, we have seen the significance of the $\sigma$ degrees of freedom.
However,  the existence of  the low-lying  scalar mesons could be 
a puzzle in QCD:
In the constituent quark model;
the meson with 
the quantum number $J^{PC}=0^{++}$  is in
the $^3P_0$ state in the nonrelativistic quark
model, which normaly implies that
the mass lies in the region from $1.2$ to $1.6$ GeV region.
So some mechanism is needed to lower the mass with 
as large as $600 \sim 800$ MeV:
(i)~ The most popular idea is the tetraquark structure proposed
by Jaffe\cite{Jaffe:1976ig}, who showed that
the color magnetic interaction between the di-quark and
the anti-di-quark
gives a large enough attraction to down the masses of 
the scalar mesons around 600 MeV.
(ii)~ Another time-honored idea is  attributing to  the possible 
collective nature of  the scalar mesons as the pion, as
the Nambu-Jona-Lasinio model\cite{Nambu:1961tp}  describes.
It is well known  that the scalar meson appears 
as a consequence of the  chiral symmetry and its dynamical
breaking as the pion does, and 
the mass of the sigma satisfies the Nambu relation,
$m_{\sigma}=2 M_f$, with $M_f$ being the dynamically generated
fermion(quark) mass, which should be valid within any Nambu-Jona-Lasinio
type model. If we put $M_f=300$ MeV, $m_{\sigma}$ becomes
600 MeV. 
It is shown that this feature essentially persists with the ${\rm U}_{\rm A}(1)$
 anomaly term incorporated\cite{Kunihiro:1987bb}.
(iii)~ The wave function of scalar mesons should  also have 
components of the meson molecule states as these states are seen
through the $\pi$-$\pi$ or $\pi$-K scattering.

The objectives of Scalar Collaboration\cite{Kunihiro:2003yj,Wada:2007cp}  are summarized as follows.
Seeing that the confidence level of the sigma meson 
and other scalar mesons has been increasing, and its physical significance 
in hadron physics and QCD is apparent,
we have been (and will be) addressing 
the following questions about the scalar mesons using lattice QCD;
 i.e., are you a pole in QCD ? i.e., 
the $\sigma$ and other low-lying scalar mesons are resonances
 in QCD or something else ?
Notice that these questions are issued as early as 2001-2002.

\section{The Scalar mesons on the Lattice;\, 
a full QCD calculation}

The Scalar Collaboration\cite{Kunihiro:2003yj} performed
a first exploratory work on the sigma in lattice QCD with dynamical quarks.
Notice that a full QCD simulation is necessary 
to properly describe the sigma with the possible contents, i.e.,
 the glueball, tetra quarks and so on.

They employed Wilson fermions and the plaquette gauge action, 
with a point source and sink.
The lattice employed is rather small, $8^3\times16$ and coarse,
$\beta = 4.8$ corresponding to a lattice spacing  $a = 0.207(9)$ fm, which
 leads to larger masses due to a mixture of higher mass states.  
In other words, the masses to be obtained in our simulation
should be considered as upper limits. 
 The following three hopping parameters are employed, i.e., $\kappa = 0.1846$, 
$0.1874$ and $0.1891$;  the  critical value is found to be
 $\kappa_c = 0.195(3)$. 

The following operator was adopted
$\hat{\sigma}(x)$$\equiv$ 
$\sum_{c=1}^3\sum_{\alpha=1}^4$
$({\bar{u}_\alpha^c(x)u_\alpha^c(x)+\bar{d}_\alpha^c(x)d_\alpha^c(x)})/{\sqrt{2}}$ ,
for creating a meson state with $I=0$ and 
$J^{PC}=0^{++}$; here
where $u$ ($d$) denotes the up-quark (down-quark) operator 
with  $c$ and $\alpha$ 
being the color and Dirac-spinor indices, respectively.
The $\sigma$ meson propagator 
consists of two terms, i.e.,
 a connected diagram and  a disconnected diagram.
Since the vacuum expectation 
value $\langle \sigma(x) \rangle$ does not vanish,
it should be subtracted from the $\sigma$ operator.
The disconnected diagram is calculated using 
the $Z_2$ noise method.
As for other  simulation parameters,
we refer to the original paper \cite{Kunihiro:2003yj} .  
 

\begin{table}[htb]
\begin{center}
Table 1:  Summary of the results. $m_{\rm con.}$ is the $\sigma$ mass estimated only from the connected
part.
\begin{tabular}{c|c|c}
\hline
\hline
$\kappa$& $m_{\sigma}/m_{\rho}$ & $m_{\rm con.}/m_{\rho}$ \\
 \hline
 0.1846        & 1.583$\pm$0.098        &  2.400 $\pm$ 0.018   \\ 
\hline
0.1874 &  1.336 $\pm$ 0.071  &  2.436 $\pm$ 0.025 \\
 \hline
0.1891 &  1.112 $\pm$ 0.060    & 2.481 $\pm$ 0.031    \\
 \hline
\end{tabular} \\
\end{center}
\end{table}

We show in Table 1 the value of $m_\sigma/m_\rho$  for each hopping parameter together
with the corresponding $m_{\rm con.}/m_{\rho}$, where $m_{\rm con.}$ denotes
the scalar meson  mass for which the disconnected diagram  is not
included. It should be noticed here that the mass
 $m_{con.}$ can be identified with the a$_0$ meson which is
the isovector meson for which the disconnected diagram does not
play any role.

\begin{center}
\begin{figure}[htb]
\rotatebox[origin=c]{-90}{
\includegraphics[width=.65 \linewidth]{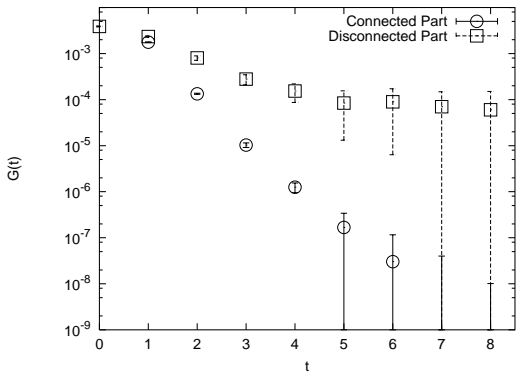}
}
\vspace{-2.0cm}
\caption{Propagators of the connected and disconnected diagrams of 
the $\sigma$ for $\kappa=0.1874$.
}
\label{fig:ConDisc}
\end{figure}
\end{center}

The individual contributions 
of the connected and disconnected parts of the $\sigma$ propagator
are shown in Fig.\ref{fig:ConDisc}, which tells us that
the connected part only shows a rapid damping 
with small error bars,
while the disconnected part overwhelming the connected part and 
dominates the $\sigma$ propagator. 
Thus,  
we see that the
$\sigma$ as a light meson results from the  disconnected 
part of the $\sigma$ propagator with  the background vacuum 
condensate subtracted.

The calculated  $m_\pi^2$, $m_\rho$, $m_\sigma$
and $2m_\pi$ 
are presented in Fig. 3 of ref.\cite{Kunihiro:2003yj},
which shows that as the chiral limit is approached, 
the $\sigma$ meson 
mass obtained from  the $\sigma$ propagator decreases
and eventually becomes smaller than the $\rho$ meson mass 
 in the chiral limit. 

\begin{center}
\begin{figure}[htb]
\includegraphics[width=1.0 \linewidth]{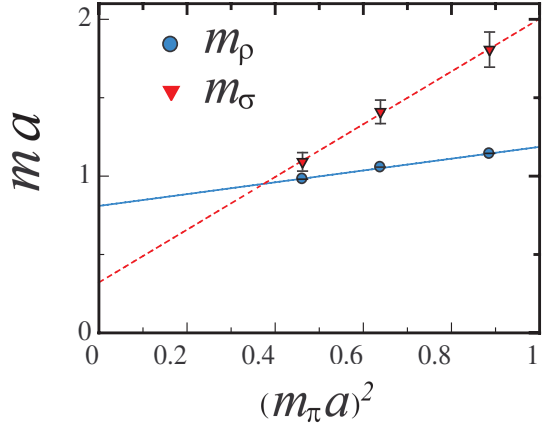}
\vspace{-1.35cm}
\caption{
The  $m_{\pi}^2$ dependence of  $m_{\rho}$ and $m_{\sigma}$
 in the lattice unit.  
Explicitly, the respective dependence is well represented
by a linear function of $m_{\pi}^2$:
$m_{\rho}a = 0.3767\times (m_{\pi}a)^2 +  0.8099$
and
$m_{\sigma}a = 1.6818\times (m_{\pi}a)^2 + 0.3219$,
respectively, with $a$ being the lattice spacing.
}
\label{fig:wada}
\end{figure}
\end{center}

It might  be more informative 
to display the masses as functions of $m_{\pi}^2$, which could be
extracted by some effective models\cite{Hanhart:2008mx}.
 The $m_{\pi}^2$ dependence of $m_{\sigma}$ and $m_{\rho}$ is
shown in Fig.\ref{fig:wada}, which is equivalent to Fig.3 of 
ref.\cite{Kunihiro:2003yj}.
 
Important points obtained in ref. \cite{Kunihiro:2003yj}
are that
the $\sigma$ propagator exhibits a pole behavior and its mass is found
to satisfy 
$m_{\pi}< m_{\sigma}\le m_{\rho}.$;
for the sigma mass to become small,
the disconnected diagram plays an essential role.
The flavored scalar meson is not light as observed experimentally;
$m_{a_0}\sim 1.9 $ GeV,  which are
 much higher than the experimental
masses, $0.6 \sim 0.8$ GeV.


\vspace{-5cm}
.\section{The $\kappa$ meson in quenched approximation}

The $\kappa$ meson is a scalar meson with the strangeness.
Recent experimental candidates
are reported \cite{Aitala:2002kr,DescotesGenon:2006uk}
to have a mass in the range from 660 to 800MeV.

The SCALAR collaboration\cite{Wada:2007cp}
 has performed a lattice QCD simulation of the $\kappa$
 has been proformed in the
quenched approximation.
The simple Wilson fermion and plaquette gauge action were
adopted for  the $20^3\times 24$ lattice,
and the lattice spacing was found to be $a=0.1038(33)$ fm.
They extracted the mass of the $\kappa$ to be $\sim$ 1.7 GeV, which is 
larger than even twice the experimental mass $\sim 800$ MeV. 
The relatively heavy mass of the $\kappa$ might
be  attributed to the absence of the disconnected diagram in 
the $\kappa$ propagator; notice that
the disconnected diagram was
essential for realizing the low-mass $\sigma$.
Indeed,  Table 1 shows that the mass of the valence 
$\sigma_v$  described solely with the connected propagator 
is far larger than the experimental value
500$-$600 MeV.  

Thus it is concluded that 
 the simple two-body  constituent-quark picture 
might  be odd with the experimentally observed $\kappa$ which has
a low mass.

\section{Summary and concluding remarks}

The $\sigma$  meson and other low-lying scalar mesons are still
a source of debates.
The understanding of the nature or the even (non-)existence is 
important for a deep understanding of the QCD vacuum 
as well as the QCD/hadron dynamics.
A full QCD lattice simulation suggests the existence of a low- lying $\sigma$ 
as a pole in QCD.
However,  its physics content, i.e.,
 a tetra quark, a hybrid with the glue ball or the q-$\bar{\rm q}$ collective state, is still obscure, 
although the fact that the disconnected diagram gives the dominant contribution
to the $\sigma$ propagator suggests that one or some of 
these exotic components are contained in the $\sigma$.
A quenched lattice calculation suggests that the kappa can not be a normal 
q-$\bar{\rm q}$ state.

There are many things to do for 
a better simulation, even apart from adopting  larger lattices
and taking a as close continuum limit as possible.
One should  make an effort to reduce errors, say, by
taking  smearing of the sources.
In fact, it is known that
the glue ball states are strongly dependent on the lattice spacing.
For studying the $\sigma$, 
one should try to reduce large errors coming from the disconnected diagrams.
The $\sigma$ has several components and above the $\pi$-$\pi$
threshold,  the variational method with multiple interpolating operators
should be adopted, where interpolation fields include
a  tetraquark operator. To see whether the obtained state is a resonance but not 
a 2$\pi$ scattering state, 
the volume dependence of the physical observables should be
calculated.  It is interesting to see the
 $N_c$-dependence of the mass and the width\cite{Pelaez:2003dy},
which will tell us a hint on the physics content of the hadron.
Maybe one should identify observables which are sensitive to the inner structure of the
$\sigma$. 
Is there any role of  the  axial anomaly\cite{Kunihiro:1987bb,Hooft:2008we}
 to realize the low-lying $\sigma$?
Exploration of the $\sigma$ at finite temperature may reveal the
nature of the particle \cite{Hatsuda:2001da}.
 Of course, it would be desirable to use chiral fermions.
The most of the above points are also true for  the $\kappa$ meson.

Finally, we refer to nice review articles for more detailed accounts on the
scalar mesons on the lattice\cite{McNeile:2007qf,Prelovsek:2008qu}

We thank Stephan Narison for inviting and giving us an opportunity
to present our work on the low-lying scalar mesons.

\end{document}